\documentclass[conference]{IEEEtran}
\makeatletter
\def\ps@headings{%
\def\@oddhead{\mbox{}\scriptsize\rightmark \hfil \thepage}%
\def\@evenhead{\scriptsize\thepage \hfil \leftmark\mbox{}}%
\def\@oddfoot{}%
\def\@evenfoot{}}
\makeatother
\usepackage{cite}
\usepackage{amsmath,amssymb,amsfonts}
\usepackage{algorithmic}
\usepackage{graphicx}
\usepackage{xcolor}
\usepackage{algorithm}
\usepackage{multirow}
\usepackage{url}
\usepackage{multicol}
\usepackage{xfrac}
\usepackage{stmaryrd}
\usepackage{pifont}
\usepackage{float}
\usepackage{subcaption}
\usepackage{listings}

\newcommand{\cmark}{\ding{51}}%
\newcommand{\xmark}{\ding{55}}%

\def\BibTeX{{\rm B\kern-.05em{\sc i\kern-.025em b}\kern-.08em
    T\kern-.1667em\lower.7ex\hbox{E}\kern-.125emX}}
\begin{document}

\title{EthClipper: A Clipboard Meddling Attack on Hardware Wallets with Address Verification Evasion}

\author{\IEEEauthorblockN{
Nikolay Ivanov, 
Qiben Yan
}
% %\\
 \IEEEauthorblockA{
 Computer Science \& Engineering, Michigan State University, East Lansing, MI, USA.}
}

\maketitle

\begin{abstract}
Hardware wallets are designed to withstand malware attacks by isolating their private keys from the cyberspace, but they are vulnerable to the attacks that fake an address stored in a clipboard. To prevent such attacks, a hardware wallet asks the user to verify the recipient address shown on the wallet display. Since crypto addresses are long sequences of random symbols, their manual verification becomes a difficult task. Consequently, many users of hardware wallets elect to verify only a few symbols in the address, and this can be exploited by an attacker. In this work, we introduce EthClipper, an attack that targets owners of hardware wallets on the Ethereum platform. EthClipper malware queries a distributed database of pre-mined accounts in order to select the address with maximum visual similarity to the original one. We design and implement a EthClipper malware, which we test on Trezor, Ledger, and KeepKey wallets. To deliver computation and storage resources for the attack, we implement a distributed service, ClipperCloud, and test it on different deployment environments. Our evaluation shows that with off-the-shelf PCs and NAS storage, an attacker would be able to mine a database capable of matching 25\% of the digits in an address to achieve a 50\% chance of finding a fitting fake address. For responsible disclosure, we have contacted the manufactures of the hardware wallets used in the attack evaluation, and they all confirm the danger of EthClipper.
\end{abstract}


\section{Introduction}\label{sec:introduction}

\begin{figure}
    \centering
    \includegraphics[width=0.9\linewidth]{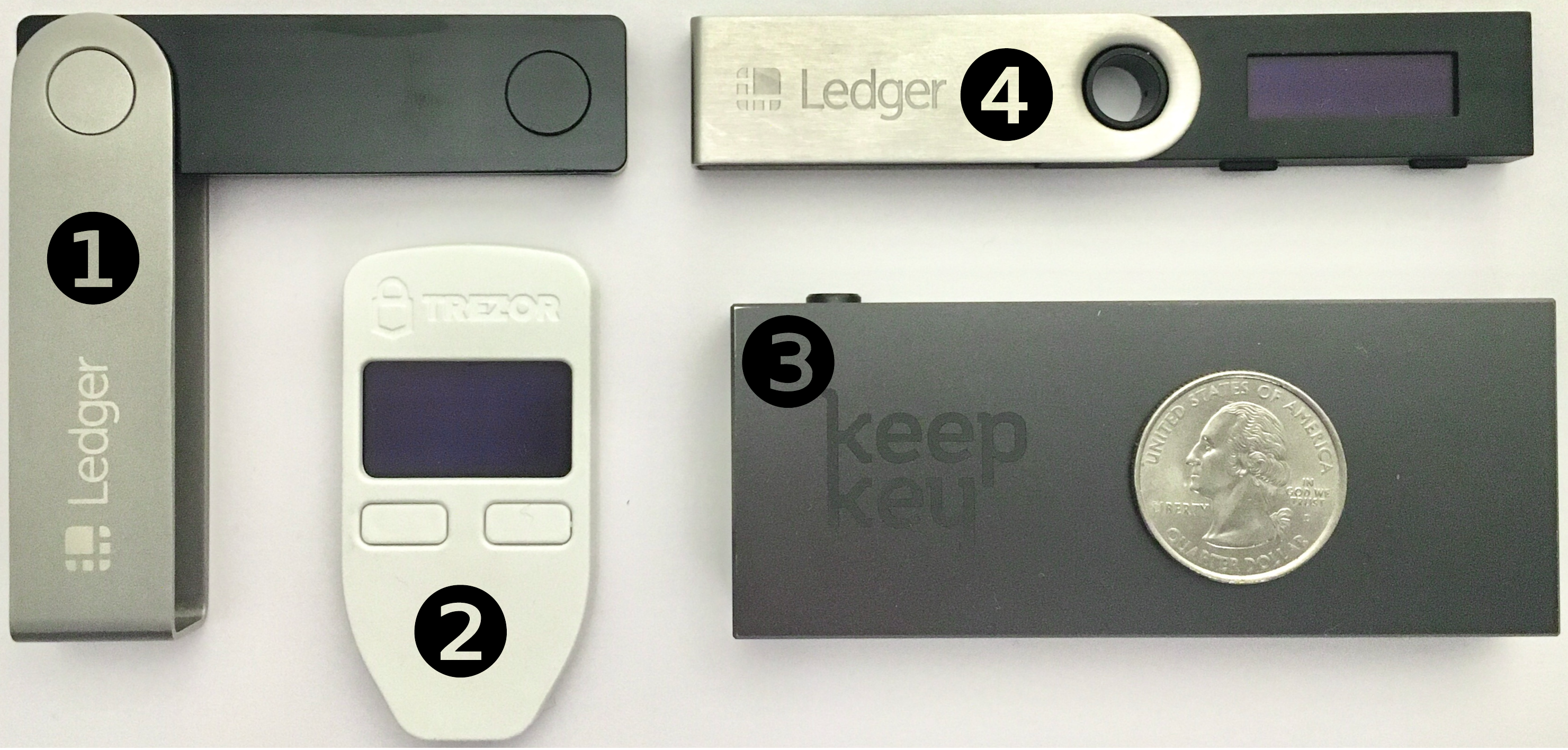}
    \caption{\textbf{Hardware wallets used in this research.} \ding{182}: \emph{Ledger Nano X}; \ding{183}: \emph{Trezor One}; \ding{184}: \emph{Keep Key}; \ding{185}: \emph{Ledger Nano S}.
    }
    \label{fig:wallets}
\end{figure}

\emph{Hardware crypto wallets}, also known as \emph{cold wallets}, are air-gapped devices that produce public-key signatures\footnote{All popular hardware crypto wallets are \emph{hierarchical deterministic} (HD) wallets~\cite{bip32-protocol}, which are capable of generating nearly infinite number of private keys (i.e., accounts) from a single secret seed.} for transactions with cryptocurrencies and smart contracts. These devices have some computing power, but they are not equipped with any networking interfaces --- to stay outside of the cyberspace. Instead, they communicate with the client computer through a secure device-to-device (D2D) channel (e.g., FIDO protocol over a USB serial bus). Hardware wallets are considered to be the most secure solution for protecting crypto funds from stealing, even in the case when the client computer is infected with malware.
Fig.~\ref{fig:wallets} shows four popular hardware wallets from three leading brands available on the market: \textit{Trezor} by SatoshiLabs s.r.o.~\cite{satoshi-labs}, \textit{Ledger Nano X} and \textit{Ledger Nano S} by Ledger SAS~\cite{ledger-sas}, and \textit{KeepKey} by ShapeShift~\cite{shapeshift}.

\begin{figure}
    \centering
    \includegraphics[width=0.8\linewidth]{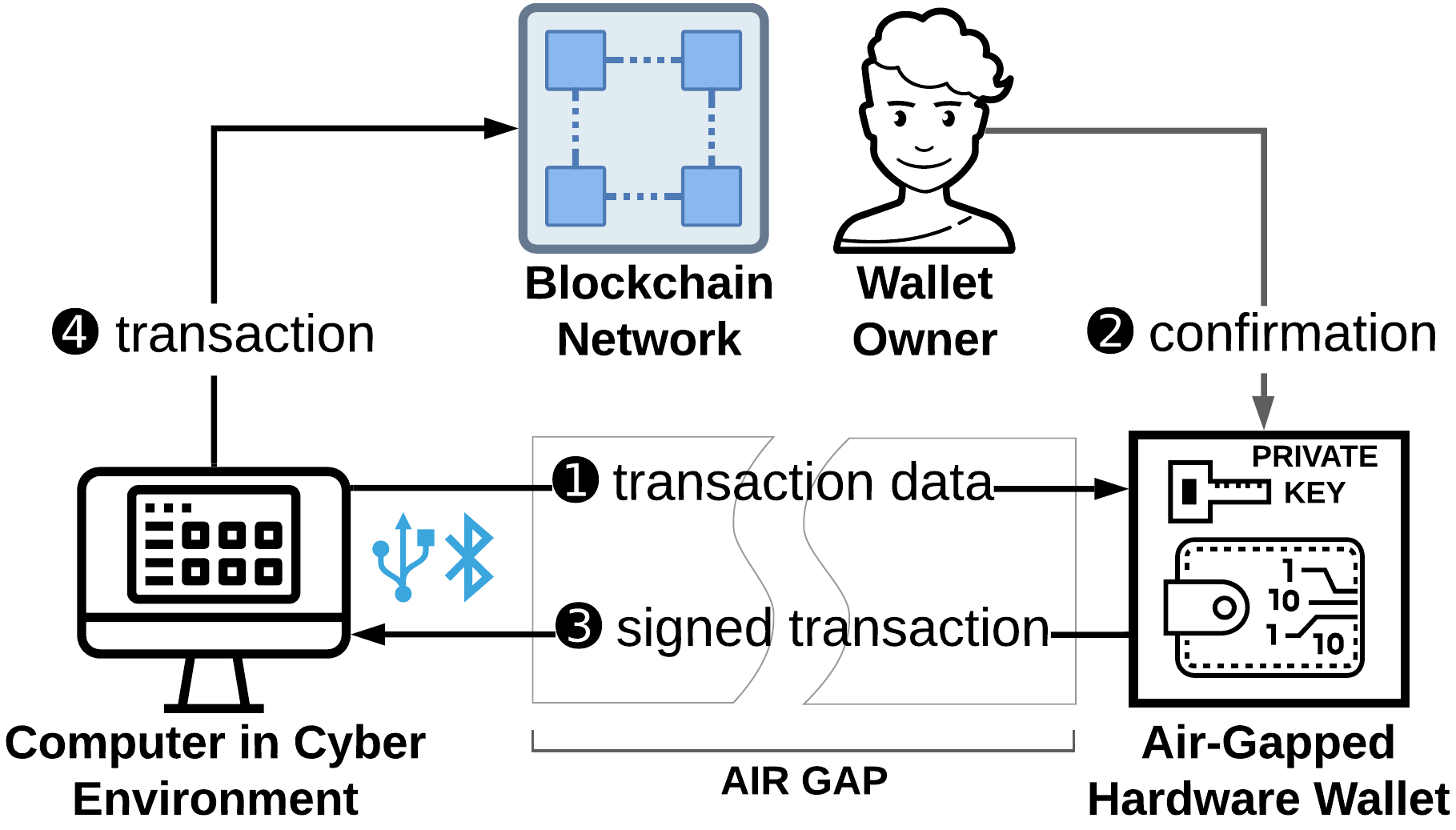}
    \caption{\textbf{General transaction workflow using a hardware wallet.} \ding{182}: The client software sends the transaction data to the hardware wallet; \ding{183}: the user verifies the data and confirms the transaction with the wallet; \ding{184}: the wallet sends the transaction signature back to the client software; \ding{185}: the client software sends the signed transaction to the blockchain network.}
    \label{fig:wallet-workflow}
\end{figure}

Fig.~\ref{fig:wallet-workflow} shows a transaction workflow with a hardware wallet. First, the client software prepares a transaction message, and sends this message over to the hardware wallet via a non-networking channel. Then, the user confirms the parameters of the transaction (such as transaction amount, recipient address, and blockchain fee) shown on the display of the wallet. After that, the wallet signs the transaction with a non-extractable private key, and sends the signature back to the client software. Finally, the client software sends the signed transaction message to the blockchain, where the transaction is executed. Unfortunately, the described chain of actions has a weak link: the attacker does not need to compromise the wallet to steal funds --- it is sufficient to tamper with the transaction data sent for signing by falsifying the address of the recipient of funds. A recent formal security analysis by Khan et al.~\cite{khan2019security} formally proves that under normal cryptographic assumptions, the user of a hardware wallet plays a crucial role in its security. One way to target the user of the hardware crypto wallet is to substitute the transaction recipient address and covertly replace it in the clipboard of the operating system.

The clipboard substitution attack, or clipboard hijacking, has been known for years~\cite{bitcoinstealer,bitcoinclip}. This attack exploits the fact that wallet users often utilize clipboard for copying a recipient address to the wallet's client app. For example, the malware called \emph{Clipsa} stole at least \$36,000 worth of Bitcoin in 2018 and 2019~\cite{clipsa}. By examining the client software provided by the three vendors of the wallets shown in Fig.~\ref{fig:wallets}, we determined that they do not discourage the use of the clipboard (e.g., by disabling the keyboard operations). In clipboard substitution attack, the malware running on the user computer detects the presence of a crypto address in the clipboard, and immediately substitutes it with another address. This attack, however, has one major weakness: the user is likely to notice the falsification of the address on the screen of the client software or on the mini-screen of the hardware wallet. Hence, our research question is: \emph{is it possible to devise a clipboard substitution attack that dodges the revelation of the address substitution during the transaction confirmation phase?}

The main insight of this work is to incorporate a social engineering component into a clipboard substitution attack. Social engineering attack techniques exploit \emph{human cognitive bias} --- an optimization mechanism of the human mind that makes conclusions based on expectation, prior experience, probability assessment, pre-existing belief, or emotions~\cite{haselton2015evolution}. One way of exploiting a cognitive bias is through \emph{visual deception}, which is actively used by attackers in email phishing via mimicking a popular website~\cite{Whittaker2010LargeScaleAC}. Another facet of cognitive bias is \emph{confirmation bias}, defined as the rejection of evidence contradicting the originally established belief~\cite{kappes2020confirmation}. We discover that both the visual deception and confirmation bias could be exploited by an attacker who tries to steal funds from the hardware wallet. Specifically, this work is inspired by our observation that hardware wallet users exhibit a strong confirmation bias about the correctness of the recipient address, resulting in a behavioral pattern to verify only several leading (or trailing) digits of a transaction address, or even skipping the verification whatsoever. \emph{The validity of this observation is confirmed by previous research~\cite{almutairi2019usability} and by manufacturers of the hardware wallets used in this research.}

In this paper, we propose a new attack called \emph{EthClipper}, which adds a social engineering component to the existing clipboard substitution technique. In \emph{EthClipper}, the attacker deploys a distributed system, called \emph{ClipperCloud}, which is used to mine and store billions of Ethereum accounts. When the malware detects an Ethereum address in the clipboard, it asks \emph{ClipperCloud} to find among the mined accounts the one that exhibits maximum visual similarity with the address in the clipboard. As a result, the visual similarity between the address on the screen and the expected address is likely to enact the victim's confirmation bias, incurring the approval of the malicious transaction. Although the \emph{Clipsa} malware~\cite{clipsa1} also attempts to match some symbols in substituted address, it uses a small address database and only targets two leading and two trailing symbols, which is very easy to reveal visually. A small human-based study conducted by Almutairi and Al-Megren~\cite{almutairi2019usability}, involving substitution of symbols in Bitcoin addresses using a \emph{KeepKey} hardware wallet, confirms that the rate of false approval of a modified address strongly correlates with the number of matching symbols. Unsurprisingly, the attacks by \emph{Clipsa} have been prevented at least 360,000 times~\cite{clipsa}. Unlike \emph{Clipsa}, \emph{EthClipper} is highly optimized to enable practical substitution of up to 25\% of the symbols in the address for achieving maximum level of deception with a limited attacker's budget, while maintaining low latency, and maximized likelihood of having a replacement address readily available.

In summary, we deliver the following contributions:

\begin{itemize}
\item We discover a new attack, called \emph{EthClipper}, against hardware crypto wallets, which combines the features of clipboard substitution, cryptographic pre-computation, and social engineering to lure the victim into confirming a transaction with a tampered recipient address.
\item We introduce a low-latency application-specific distributed system, called \emph{ClipperCloud}, that performs computation and storage needed for the \emph{EthClipper} attack outside of the victim's computer.  This makes the attack a realistic one, which is easy to carry out.
\item We implement \emph{EthClipper} malware, and test it using four popular hardware wallets from three manufacturers.
\item We implement the \emph{ClipperCloud} system and test it on four server deployments. Our evaluation shows that \emph{ClipperCloud} exhibits a low query latency, and \emph{EthClipper} attack adapts to a flexible range of setups and budgets.
\item For responsible disclosure, we have communicated the details of the attack to the manufacturers of the wallets used in this research and received the confirmation from all of them that the attack is potentially dangerous.
\end{itemize}

\section{EthClipper Attack Design and Analysis}\label{sec:design}
In this section, we elaborate on the technical details of the \emph{EthClipper} attack, and then describe the workings of the \emph{EthClipper} malware, followed by elaboration on the \emph{ClipperCloud} system needed for the attack.

\subsection{Attack Overview}

\begin{figure}
    \centering
    \includegraphics[width=0.6\linewidth]{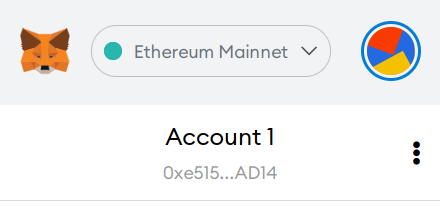}
    \caption{\textbf{Replacing 80\% of address symbols with an ellipsis in MetaMask, one of the most popular Ethereum wallets.}}
    \label{fig:metamask}
\end{figure}

\begin{figure}
    \centering
    \includegraphics[width=\linewidth]{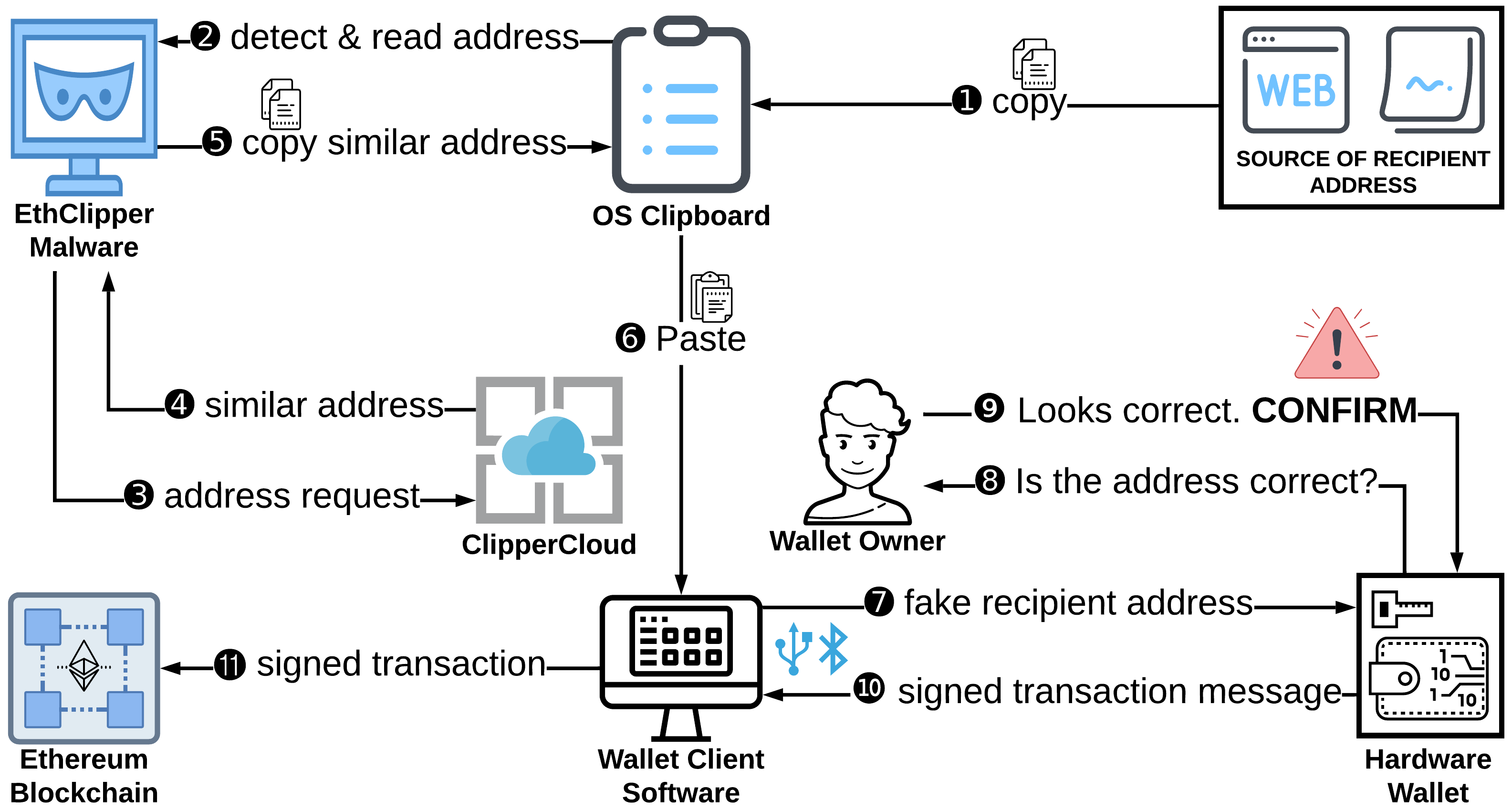}
    \caption{\textbf{Workflow of the \emph{EthClipper} attack.} \ding{182}: The owner of the wallet copies a recipient address to the clipboard from the source (e.g., website); \ding{183}: the \emph{EthClipper} malware detects the address in the clipboard; \ding{184}: the malware connects to \emph{ClipperCloud} to request an address that is similar to the one in the clipboard; \ding{185}: \emph{ClipperCloud} replies with a similar address; \ding{186}: \emph{EthClipper} malware places the substitute address from \emph{ClipperCloud} to the clipboard; \ding{187}: the user of the wallet pastes the address from the clipboard to the hardware wallet's client software; \ding{188}: the client software sends the transaction data, which includes the replaced (fake) recipient address, to the hardware wallet for signing; \ding{189}: the hardware wallet asks the user to confirm the parameters of the transaction (by pushing a button on the wallet); \ding{190}: the user of the hardware wallet, who is prone to a confirmation bias, confirms the transaction without verifying all of the symbols of the recipient address; \ding{191}: the wallet signs the transaction using the air-gapped private key and sends the signature to the wallet's client software; \includegraphics[width=0.27cm]{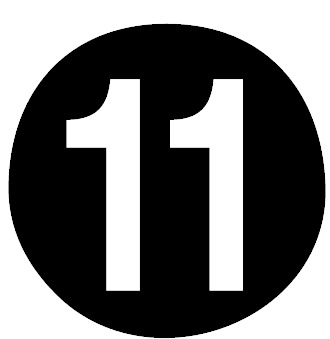}: finally, the wallet client software sends the signed transaction to the Ethereum blockchain, where the transaction is executed.}
    \label{fig:rust-attack}
\end{figure}

The manufacturers of popular hardware crypto wallets openly state that hardware wallets are the best devices for storing crypto accounts. Yet, the \emph{EthClipper} attack bypasses the air-gapped protection of hardware wallets in order to steal money from the user's Ethereum account by falsifying the transaction recipient address with an address belonging to the attacker. The \emph{EthClipper} attack uses the clipboard hijacking technique to falsify the data sent to a hardware wallet without compromising the wallet itself. However, unlike previous attacks of this kind, \emph{EthClipper} makes it harder for the user to recognize a falsification.  Fig.~\ref{fig:rust-attack} shows a general workflow of the attack. The attacker infects the victim's computer with malware using one of a plethora of available techniques~\cite{chakkaravarthy2019survey,sahay2020evolution}. The malware monitors the clipboard of the user account for appearance of a valid public crypto address. Once the address is discovered in the clipboard, the malware immediately contacts the pre-deployed distributed system, called \emph{ClipperCloud}, which stores a database of similar addresses that have been mined in advance (see Section~\ref{sec:clippercloud} for details). After receiving the matching visually similar address from \emph{ClipperCloud}, the malware replaces the original address in the clipboard with the forged one.

\begin{figure}
    \centering
    \includegraphics[width=\linewidth]{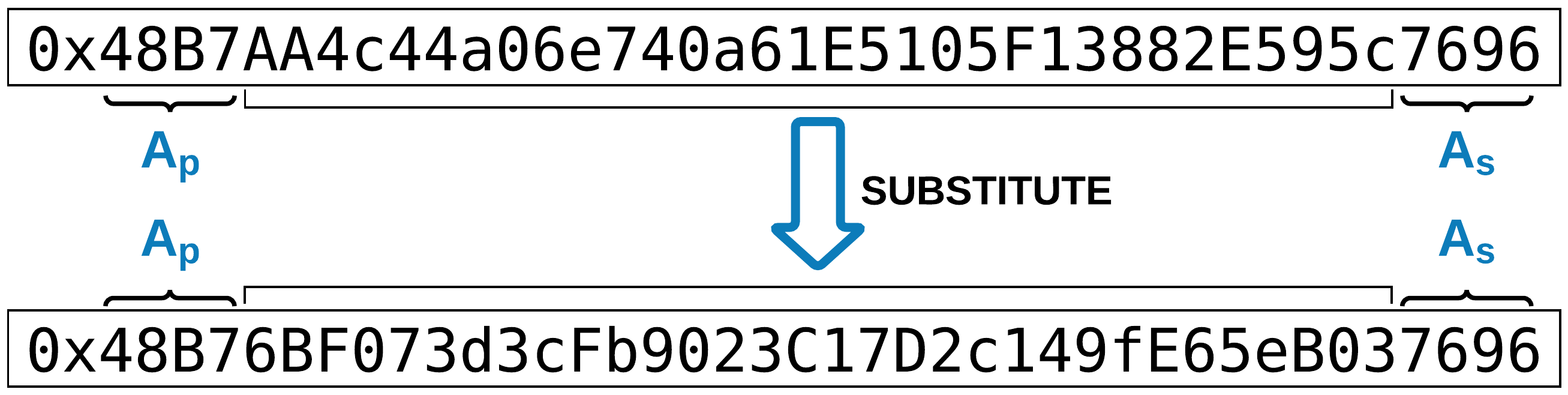}
    \caption{\textbf{Address substitution pattern.} The substituted address has the same number of matching prefix and suffix symbols (or one more in the prefix, when the number of symbols is odd), i.e, $\lceil\frac{N}{2}\rceil$ in the prefix $A_p$, $\lfloor\frac{N}{2}\rfloor$ in the suffix $A_s$, $N$ total. When verifying the address, many users check only a few symbols in the prefix, and sometimes a few symbols in the suffix.}
    \label{fig:address-substitution}
\end{figure}

Our observation suggests that users of hardware wallets tend to verify a few leading \emph{and} trailing symbols in the address. Moreover, many popular Ethereum wallets, such as MetaMask, indirectly suggest the normality of skipping the internal symbols of the address by incorporating this feature in the user interface (see Fig. \ref{fig:metamask}). This observation is independently confirmed by three manufacturers of hardware wallets, and leads us to the design of the attack that substitutes the address with matching $\lceil\frac{N}{2}\rceil$ symbols in the prefix and $\lfloor\frac{N}{2}\rfloor$ symbols in the suffix (see Fig.~\ref{fig:address-substitution}). Moreover, our observation of Ethereum address checking by hardware wallet users reveals the habit of not verifying more than 4 symbols in the prefix and 4 symbols in the suffix, suggesting that $N=8$ is likely to be sufficient amount of matching symbols in many cases. Since the attack is opportunistic, there is no need for the attacker to succeed every time. However, the probability of success is obviously growing with larger values of $N$. Furthermore, a large amount of funds involved in a cryptocurrency transaction does not necessarily entail increased vigilance by the user. For example, in 2016, an attack on an Ethereum smart contract, known as the DAO attack, incurred a damage worth approximately \$50 million due to a simple reentrancy vulnerability, which had been known and well-researched for years prior to the attack~\cite{cecchetti2021compositional} --- despite high amounts of money at stake, none of the investors in the infamous smart contract was able to notice the bug that was exploited in the attack. Therefore, we do not exclude users of large transactions from the scope of potential victims of \emph{EthClipper}.

\begin{figure}[h]
\captionsetup[subfigure]{justification=centering}
\centering
\begin{subfigure}{.24\textwidth}
  \centering
  \includegraphics[width=1.3in]{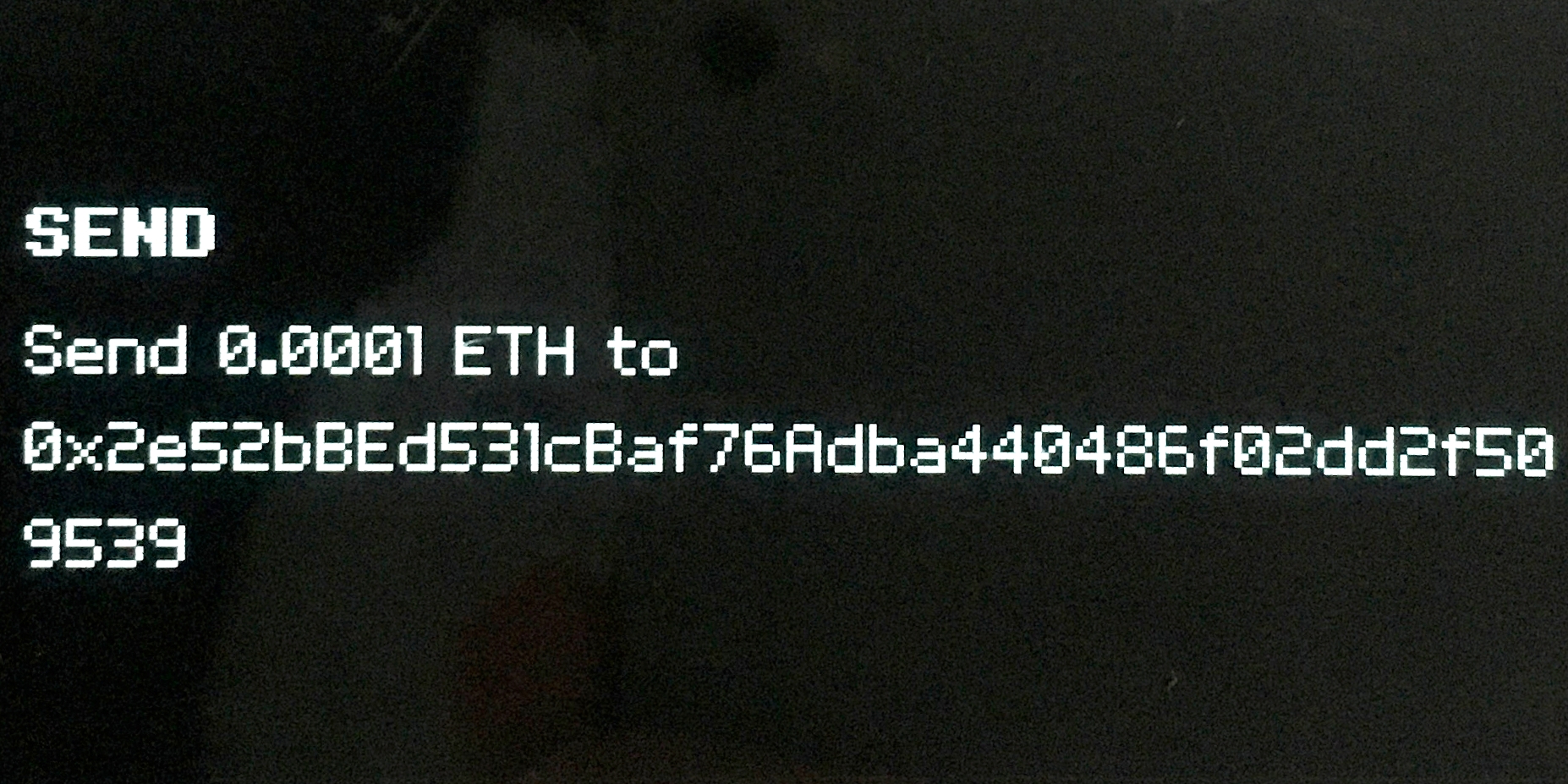}
  \vspace{-5pt}
  \caption{\emph{KeepKey}}
  \label{fig:keepkey}
\end{subfigure}~~~
\begin{subfigure}{.24\textwidth}
  \centering
  \includegraphics[width=1.3in]{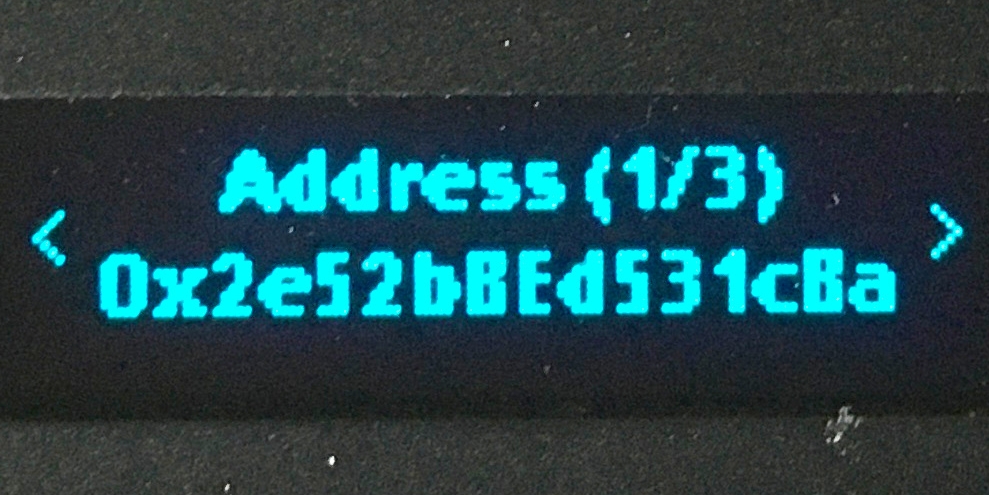}
  \vspace{-5pt}
  \caption{\emph{Ledger Nano S}}
  \label{fig:nanos}
\end{subfigure}

\vspace{0.15cm}

\begin{subfigure}{.24\textwidth}
  \centering
  \includegraphics[width=1.3in]{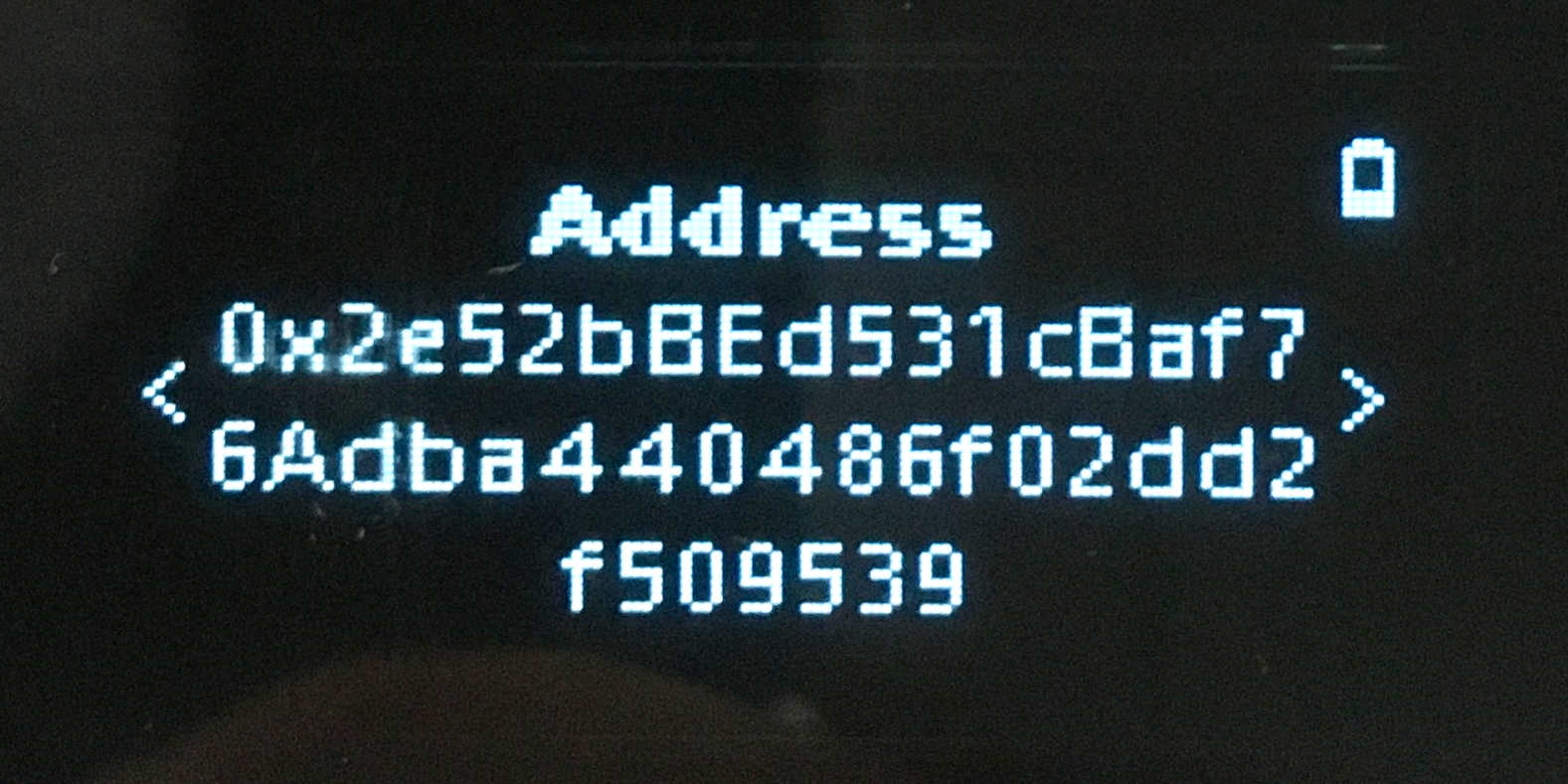}
  \vspace{-5pt}
  \caption{\emph{Ledger Nano X}}
  \label{fig:nanox}
\end{subfigure}~~~
\begin{subfigure}{.24\textwidth}
  \centering
  \includegraphics[width=1.3in]{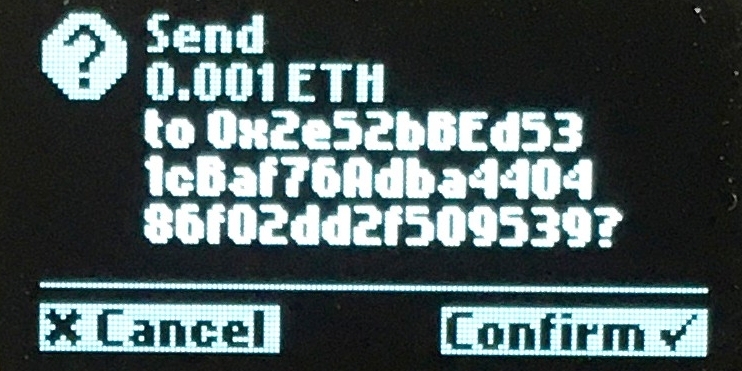}
  \vspace{-5pt}
  \caption{\emph{Trezor One}}
  \label{fig:trezor}
\end{subfigure}

\caption{\textbf{Ethereum cryptocurrency transaction confirmation in popular hardware wallets.}
}
\label{fig:screens}
\end{figure}

When the user pastes the address in the hardware client application, he/she has to confirm the parameters of the transaction on the screen of the wallet, which includes the recipient address, as shown in Fig.~\ref{fig:screens}. Since the address on the screen is visually similar to the expected one (i.e., the two addresses have matching prefixes and suffixes), the victim might fail to notice the substitution. Our informal communications with several users of hardware wallets confirm that most of them, when verifying the recipient address, examine only first and last several digits, or none at all. Finally, the user pushes the confirmation button on the hardware wallet and sends the funds to the address which corresponding private key is stored on \emph{ClipperCloud}, and therefore known to the attacker. \emph{EthClipper} is optimized for the specifics of Ethereum, which allows for the attacker to maximize the social engineering effect of the attack, which existing malware, such as \emph{Clipsa}, fails to achieve. However, it is possible to independently develop a similar malware and associated distributed service optimized for other formats of addresses, such as Bitcoin.

\subsection{EthClipper Malware}
In this research, we design a malware that allows to bypass the air-gapped protection of a hardware wallet through the \emph{EthClipper} attack, which uses clipboard substitution as a carrier. \emph{EthClipper} malware is a program that persistently runs on the background, monitoring the clipboard of the current user. An important feature of \emph{EthClipper} malware is that it does not require any special user privileges or hardware access. Moreover, it can be implemented as a cross-platform Python or Node.js script. Once the malware detects an Ethereum address in the clipboard, it immediately submits a UDP request to the \emph{ClipperCloud} system, which replies with a substitute address, if one is found. As soon as the substitute address is received, the malware injects it in the clipboard. Intuitively, it is very important for the malware to substitute the address very quickly, before the user pastes the address to the wallet client application. 
The manufacturers of the hardware wallets used in this study confirmed for us that currently there is no defense against \emph{EthClipper} attack. Thus, given the decentralized nature of the Ethereum blockchain, if the attack is deployed and subsequently revealed by one or multiple users, it would require an extensive publicity and substantial amount of time to alert all potential victims of the attack. Next, we elaborate on the architecture of \emph{ClipperCloud}, which provides the storage model that allows to achieve a low response latency to ensure the success of the attack.

\subsection{ClipperCloud}\label{sec:clippercloud}

In order to make \emph{EthClipper} practical for a real-world attacker, the abundant storage and heavy computation needed for the attack must be outsourced to a distributed service. \emph{ClipperCloud} is a distributed system that has two main purposes: it mines malicious addresses for the attacker, and it stores these addresses in a way that allows to query them very quickly. Next, we elaborate on the architecture, computation, and storage model of \emph{ClipperCloud}.

\subsubsection{ClipperCloud Architecture}

The \emph{EthClipper} attack requires a heavy computation (for mining similar addresses), as well as a large storage (for keeping the pre-mined addresses ready for the malware and storing their corresponding private keys for the attacker to withdraw stolen money). Moreover, to make \emph{ClipperCloud} suitable for the \emph{EthClipper} attack, the system must meet the following four major requirements. First, regardless of the size of the database, the system must respond to the malware requests very quickly, in order to replace the recipient address in the clipboard before the user pastes it. Second, the computation and storage may need to be split between multiple servers because a single server might not have sufficient resources required for the attack. Third, the \emph{EthClipper} malware is likely to have multiple instances, so the \emph{ClipperCloud} system must be able to serve them all. Fourth, the system must be flexible enough to support adding additional computation and storage, as well being capable for easy reconfiguration after the address database is fully mined. To satisfy these desiderata, we design \emph{ClipperCloud} in a way that it can split resources across multiple servers. To achieve that goal, each server performs communication with malware, computation and storage in three parallel processes.

\begin{figure}
    \centering
    \includegraphics[width=0.9\linewidth]{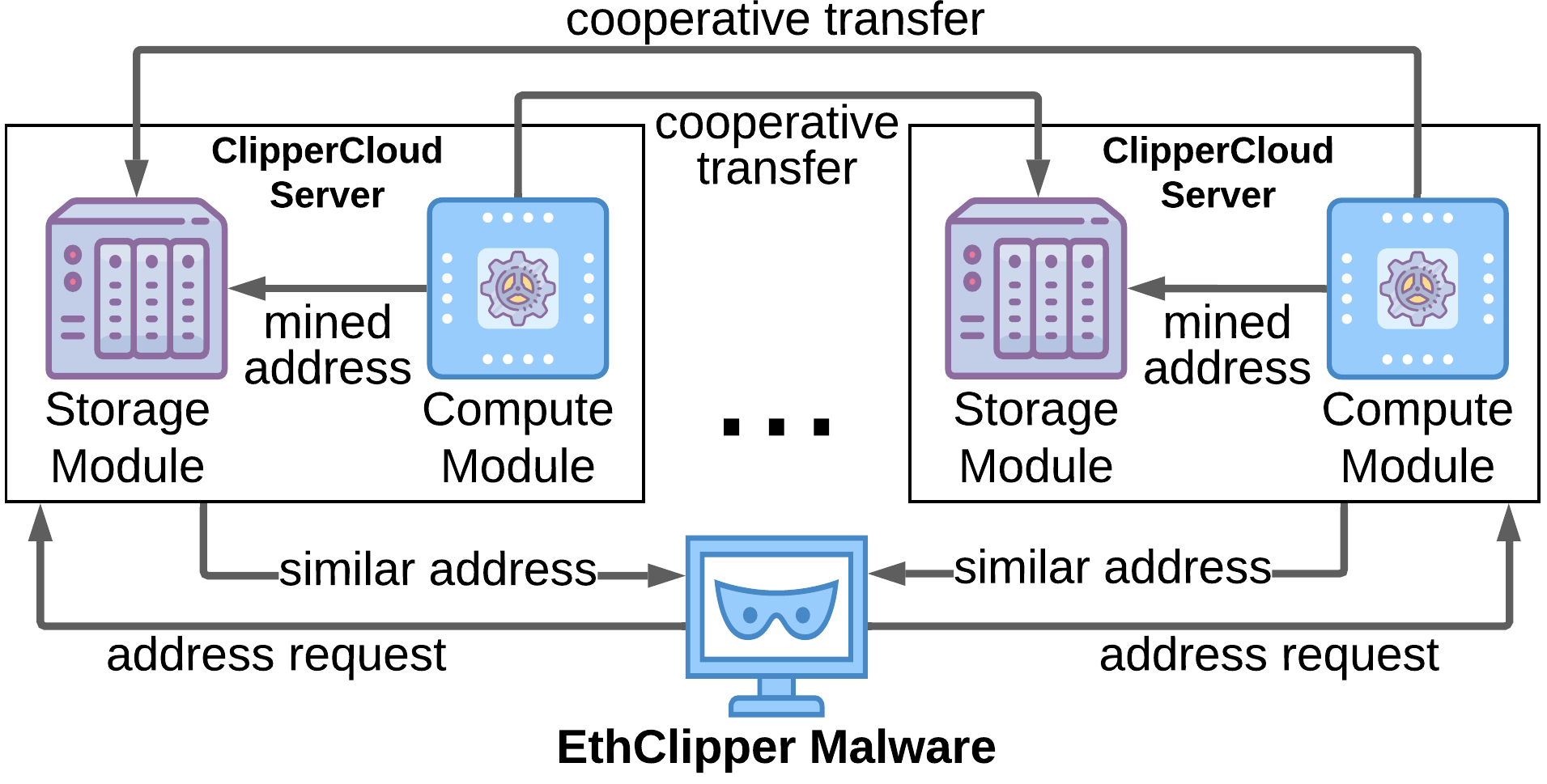}
    \caption{\textbf{\emph{ClipperCloud} workflow.}
    }
    \label{fig:chads-arch}
\end{figure}

Fig.~\ref{fig:chads-arch} shows the basic architecture of \emph{ClipperCloud}. The distributed system can have one or several servers. Each server has a compute module, which performs address mining. It also has a storage module, which saves mined addresses, along with their corresponding private keys. Each server is responsible for storing addresses corresponding to a certain range of matching symbols, while the compute module can produce addresses for any of the servers (because the result of the random guessing is unpredictable). If the compute module on one server finds a matching address for another server, it stores the result in a temporary buffer. When the buffer is full, the server transfers these addresses over to the corresponding server --- this procedure is called the \emph{cooperative transfer}. 

We conducted a preliminary testing using a high-performance Microsoft Azure H-Series server with 60 CPUs, which revealed that the cooperative transfer overhead was between 200 and 300 megabytes per minute, while the available bandwidth is normally 1 Gbps in the uplink direction and 9 Gbps in the downlink direction. This result confirms that there is no risk of a traffic bottleneck incurred by the cooperative transfer. Also, we experimentally confirm that despite the increased network traffic, the cooperative transfer delivers at least 50\% faster database population compared to discarding out-of-range addresses --- we attribute this phenomenon to the benefits of the usage of direct memory access (DMA) or similar hardware extensions by the servers, which allow to perform disk operations with minimal CPU involvement.

\subsubsection{Compute Module and Address Mining}
In order to conduct the \emph{EthClipper} attack, the attacker needs to have a large database of Ethereum accounts readily available for address substitution. In this work, we call the process of population of such a database the \emph{address mining}, which is performed by the \emph{ClipperCloud} module called the \emph{address miner}. \emph{ClipperCloud} address miner is a multi-threaded program that generates random Ethereum accounts. For example, consider the address substitution depicted in Fig.~\ref{fig:address-substitution}. The bottom address in this figure will be stored in the slot 48B77696 (in hexadecimal representation) within the \emph{ClipperCloud} address space. Each address in the \emph{ClipperCloud} address space translates into an absolute address within one of the \emph{ClipperCloud} servers. The address miner, when a new account is created\footnote{By account, we assume a pair made of a private key and corresponding public address. In Ethereum, the address of an account is calculated as a 160-bit prefix of the Keccak256 hash of the account's public key.}, either forwards the account to the storage module, or eventually sends this account to the \emph{ClipperCloud} server where it belongs (as part of the cooperative transfer). If there is already an account stored in that slot, the new account is ignored.

\subsubsection{Fixed-Field Storage}

The account records produced by the address miner should be stored at \emph{ClipperCloud} in a way that any requested address substitute must be found very quickly --- otherwise, the malware will not be able to substitute the address in the victim's clipboard within the short period of time between copying and pasting of the address. To guarantee instant response to a record search, \emph{ClipperCloud} stores records as hexadecimal strings in a fixed-field database, so its total storage requirement $S_{tot}$ can be calculated as $S_{tot} = (S_{prk} + S_{pa}) \times 16^{N}$, where $S_{prk}$ is the size of a private key, $S_{pa}$ is the size of a public address, and $N$ is the number of matching symbols (both prefix and suffix). Since \emph{EthClipper} targets only Ethereum users, $S_{prk}$ and $S_{pa}$ can be replaced with their respective numerical values of 64 and 40 bytes, i.e., $S_{tot} = 104 \times 16^{N}$.

\begin{figure}
    \centering
    \includegraphics[width=\linewidth]{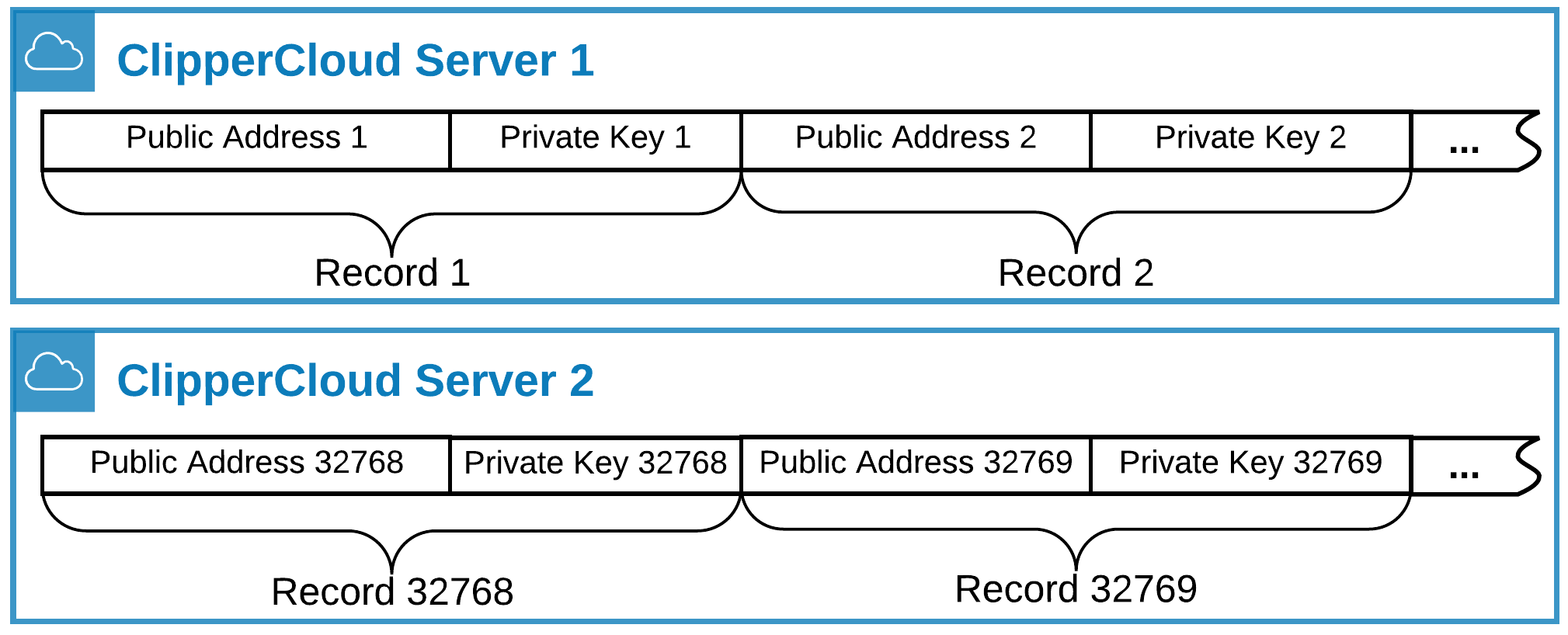}
    \caption{\textbf{Overview of the \emph{ClipperCloud} storage format.}
    }
    \label{fig:storage-format}
\end{figure}

As shown in Fig.~\ref{fig:storage-format}, the records in \emph{ClipperCloud} are stored sequentially in fixed-sized fields. The length of one field is 104 bytes (40-byte address concatenated with 64-byte private key). This allows to access the records with the time complexity in the order of $\mathcal{O}(1)$. To access the record within the file storage, the server needs to perform a single \texttt{lseek}\footnote{\texttt{lseek} is a system call in POSIX-compatible operating systems (e.g., Linux) that moves the read/write position (called \emph{offset}) within a file. This operation is intended to have a constant-time complexity.} operation within the data file with the offset set as $104 \times ([A_p \baro A_s] - a_0)$, where $[A_p \baro A_s]$ is the number resulting from the concatenation of the prefix $A_p$ and the suffix $A_s$; $a_0$ is the first value in the range of record numbers assigned to the current \emph{ClipperCloud} server. Additionally, \emph{ClipperCloud} allocates storage for cooperative transfer buffers, as well as a little space for logging successful requests (in order to inform the attacker which accounts have stolen funds). Both of these additional storage components remain constant and much smaller than the storage of records with substantially large $N$, so we exclude these insignificant values from the storage analysis.

\subsection{Address Mining Analysis}\label{sec:compute}

Each newly generated random address might match a previously stored \emph{ClipperCloud} record. Moreover, the more addresses \emph{ClipperCloud} generates, the higher the probability of collision with an already stored record, which slows down the rate of adding new records to the database of similar addresses. Since the \emph{EthClipper} attack is opportunistic by its nature, let us assert that 95\% coverage of available similar addresses is satisfactory for an attacker. In other words, we assume that the \emph{ClipperCloud} database is fully-mined if any given address request has a 95\% probability of success. Here, we deliver a formal argumentation regarding the compute complexity of the brute-force similar address mining that is probabilistically necessary for achieving the 95\% target coverage.

\noindent\textbf{Claim 1:}
\emph{A random set of $3 \cdot M$ integer numbers from the interval $[0,M-1]$ is expected to have at least $0.95 \cdot M$ distinct values.}

\noindent\textbf{Proof:} Let us consider a set $S$ of properly random numbers $S_i \in \mathbb{Z}$, $ 0 \le S_i \le M-1$, and $|S| = 3M$. Each number in the set is expected to have a certain probability of collision with at least one other number in the set, i.e.:
\begin{align}\label{eq:collision}
    p = Pr(\exists i \in [1,3M] \: \exists j \in [1,3M] : i \neq j \wedge S_i = S_j)
\end{align}
The expected value of $p$ is consistent with the well-known \emph{birthday paradox}~\cite{katz2020introduction}, in which the expected proportion of resulting distinct values $C$ of $m$ possible values in the random sample of size $n$ can be determined by the Taylor's approximation, shown below, that delivers a provably narrow margin of error~\cite{sayrafiezadeh1994birthday}:
\begin{align}\label{eq:taylor}
    C = 1 - e^{-\sfrac{n}{m}}
\end{align}
Next, let us apply Eq.~(\ref{eq:taylor}) towards the constraints described in Eq.~(\ref{eq:collision}):
\begin{align}\label{eq:t95}
    C = 1 - e^{-\sfrac{3M}{M}} = 1 - \frac{1}{e^3} \approx 0.95021
\end{align}
Therefore, the expected number of distinct values in the set of $3M$ random integer numbers between $0$ and $M-1$ is at least 95\% of total possible distinct numbers, i.e., $C \ge 0.95 \cdot M$. $\blacksquare$

\noindent\textbf{Corollary of Claim 1:}
\emph{In order to attain the coverage of at least 95\% of similar addresses, the attacker is expected to generate $3 \cdot 16^N$ random Ethereum accounts.}

\noindent\textbf{Proof:}
An Ethereum address is a 160-bit prefix of a Keccak256 digest of the public key of the account, which is derived from the 256-bit random private key of the account via the \emph{secp256k1} elliptic curve algorithm~\cite{wood2014ethereum}. Assuming that any hexadecimal digit position of an Ethereum address expresses an equal probability of its 16 possible values (i.e, $0$ through $F$), then any subset of $N$ digits in a random address is essentially an integer number from the interval $[0,16^N-1]$. Therefore, Claim~1 can be applied to the address mining by \emph{ClipperCloud}, with $M=16^N$. Consequently, in order for the attacker to achieve a minimum 95\% coverage of similar addresses with $N$ matching digits, $3 \cdot 16^N$ random Ethereum accounts must be generated. $\blacksquare$

For the purpose of generality, let us denote the multiplier 3 in \emph{Claim 1} as $\tau$ (i.e., $\tau = 3$). Following the same logic, we can leverage different target coverage values by changing $\tau$. For example, when $\tau = 1$, we may expect at least 63\% of database coverage, i.e.:
\begin{align}
        C = 1 - e^{-\sfrac{1M}{M}} = 1 - \frac{1}{e} \approx 0.6321
\end{align}

Similarly, when $\tau = 0.7$, the approximate coverage is 50\%, which means that the attacker needs to generate $0.7 \cdot 16^N$ accounts to achieve 50\% probability of successful $N$-digit match for a given random address. To confirm the correctness of the above argumentation, we conducted a small experiment for the case of 95\% coverage: we generate 3 million random numbers between 0 and 999,999, % in Python, 
and add them to the set that prohibits duplicates. The resulting size of the set was 950,188, which is consistent with Eq.~(\ref{eq:t95}).

\section{Implementation and Evaluation}\label{sec:evaluation}

\subsection{Implementation}
In order to demonstrate that \emph{EthClipper} is feasible, we implement it and perform a thorough testing of its parameters using four different hardware wallets from three manufacturers. We implement our \emph{EthClipper} malware prototype using Python 3.7.5 with \texttt{socket} and \texttt{clipboard} libraries. The \emph{ClipperCloud} prototype is implemented using Node.js JavaScript 10.15.2 with \texttt{dgram}, \texttt{buffer}, \texttt{fs}, and \texttt{Web3.js} libraries. After the manufacturers of the hardware wallets deploy the defense, we intend to publish the source code of our implementation under an open-source license for testing, reproduction, independent evaluation, and follow-up research.

We test our implementation using four hardware wallets: \emph{Ledger Nano X}, \emph{Trezor One}, \emph{KeepKey}, and \emph{Ledger Nano S}. \emph{Ledger Nano X} supports both Bluetooth and USB connections, but we use only USB, for fair comparison. For \emph{Trezor One} and \emph{KeepKey}, we use the vendor's bridge software installed on Ubuntu 20.04, and the vendors' web apps (Trezor Ethereum Wallet and ShapeShift) in Google Chrome web browser. For the Ledger wallets, we use the vendor's bridge software and the vendor-provided cross-platform desktop GUI application. Then we execute the workflow of the attack three times with each of the four wallets, confirming that the attack executes as expected and that the similarity of the addresses shown for confirmation on the screen of the wallets indeed have a deceptive quality on the human cognition.

\subsection{Storage Requirement}

\emph{EthClipper} can be used with a wide spectrum of \emph{ClipperCloud} configurations, thereby leveraging the balance between the number of matching symbols, address mining time, address database coverage, and the budget of the attacker. Table~\ref{tab:storage-requirement} shows some possible \emph{ClipperCloud} storage configurations that the attacker may use. As we can see from the table, if the attacker wants to match only 4 symbols in the address, \emph{ClipperCloud} needs to store about 6.5 megabytes of information. However, in order to match 11 symbols, the storage requirement increases to over 1.6 petabytes, which would require about 820 2-terabyte hard drives, unattainable by most attackers. The storage configurations for up to 9 matching symbols are easily attainable with retail storage devices or affordable cloud solutions. The database matching 10 symbols, requiring 104 Tb, is also achievable with relatively affordable retail options. For example, as of mid-April 2021, two WD EX4100 56TB off-the-shelf network access storage (NAS) units can provide the attacker with sufficient memory for addresses with 10 matching symbols at a total cost of under \$4,200. Thus, we assume that the attacker's \emph{ClipperCloud} database has the maximum capability for replacing 10 symbols, which is 25\% of the total Ethereum address length. Unlike \emph{Clipsa}, \emph{EthClipper} allows to match larger number of symbols in the address, thereby substantially increasing the odds of success.

\begin{table}
    \centering
    \caption{\textbf{Cumulative address storage requirement.}}
    \label{tab:storage-requirement}
    \begin{tabular}{|c|c|c|c|c|c|}
        \hline
         \textbf{\scriptsize Matching} & \multirow{2}{*}{\emph{\textbf{\scriptsize EthClipper}}} & \multirow{2}{*}{\emph{\scriptsize Clipsa}} &\multicolumn{3}{c|}{\textbf{\small Storage requirement per server}} \\
         \cline{4-6}
         \textbf{\scriptsize symbols} & & & 1 server & 5 servers & 10 servers \\
         \hline \hline
         4 & \cmark & \cmark & 6.5 Mb & 1.6 Mb & 665.6 Kb\\
         \hline
         5 & \cmark & \xmark & 104 Mb & 20.8 Mb & 10.4 Mb \\
         \hline
         6 & \cmark & \xmark & 1.625 Gb & 332.8 Mb & 166.4 Mb \\
         \hline
         7 & \cmark & \xmark & 26 Gb & 5.2 Gb & 2.6 Gb \\
         \hline
         8 & \cmark & \xmark & 416 Gb & 83.2 Gb & 41.6 Gb \\
         \hline
         9 & \cmark & \xmark & 6.5 Tb & 1.3 Tb & 665.6 Gb \\
         \hline
         10 & \cmark & \xmark & 104 Tb & 20.8 Tb & 10.4 Tb \\
         \hline
         11 & \cmark & \xmark & 1.625 PB & 332.8 Tb & 166.4 Tb \\
         \hline
        %  ALL & $1.38 \times 10^{38}$ Gb \\
        %  \hline
    \end{tabular}
\end{table}

\subsection{Query Latency Evaluation}

\begin{figure}
    \centering
    \begin{subfigure}{.24\textwidth}
  \centering
  \includegraphics[height=1.1in]{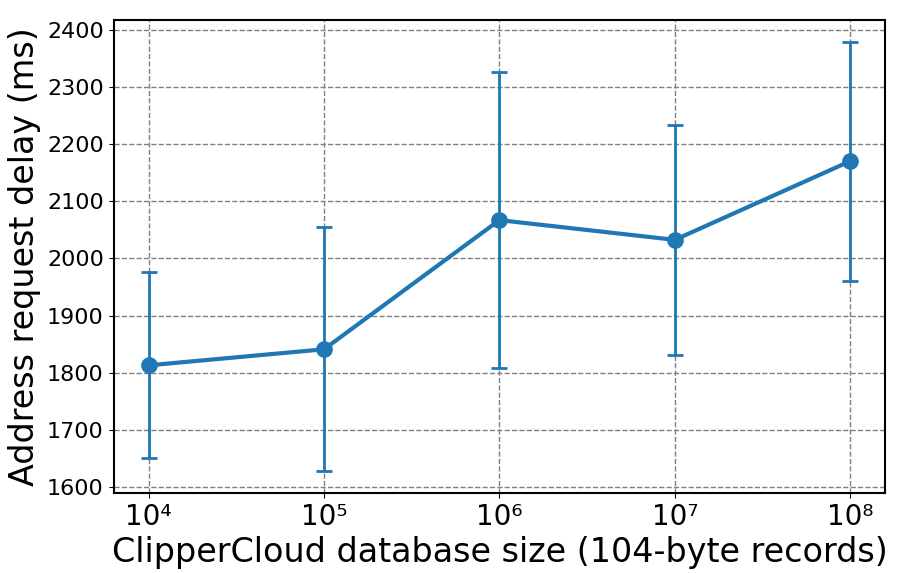}
  \caption{\emph{Address request.}}
  \label{fig:eval1}
\end{subfigure}~
\begin{subfigure}{.24\textwidth}
  \centering
  \includegraphics[height=1.1in]{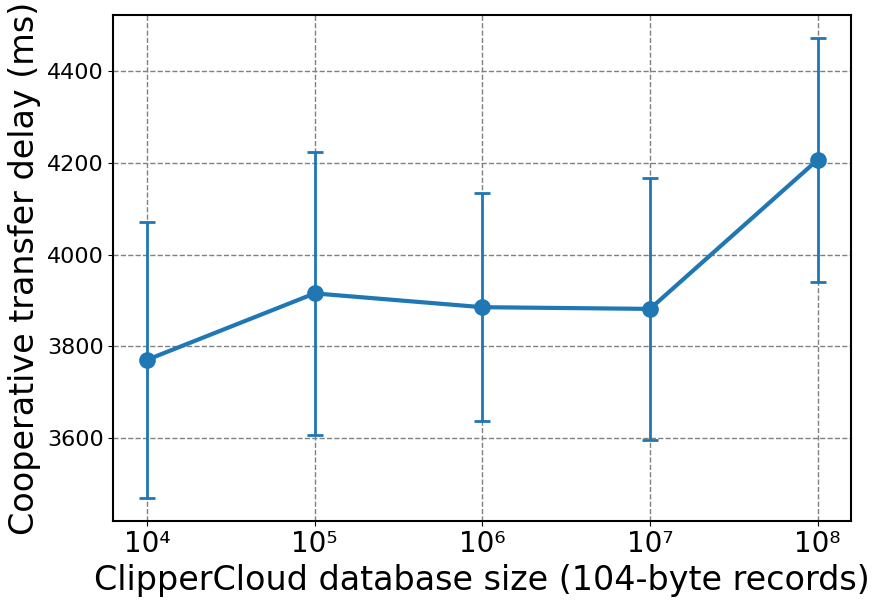}
  \caption{\emph{Mined account transfer.}}
  \label{fig:eval2}
\end{subfigure}

    \caption{\textbf{Average transmission delay.}}
    \label{fig:eval}
\end{figure}

The delay between the address request submitted by \emph{EthClipper} malware and the response by \emph{EthCloud}, denoted the \emph{query latency}, is crucial for the success of the attack because the address has to be replaced with the similar one before the user pastes the address to the wallet client. In order to evaluate the delay of similar address requests, we conduct 5 experiments, each including 20 measurements (100 measurements total). For each experiment, we exponentially increase the number of addresses stored in one \emph{ClipperCloud} server from $10^4$ to $10^8$. Then, for each of the 5 experiments, we measure the delay of requesting a similar Ethereum address in milliseconds. We use two \emph{ClipperCloud} servers (one in San Francisco, another one in New York) using DigitalOcean Droplet service, both with the following configuration: CPU-Optimized 32-CPU servers with 400 Gb SSD, 64 Gb RAM, running Ubuntu 20.04 LTS x64. We test three different attacker's Internet connection types: 100 Mbps home cable modem, 60 Mbps home Wi-Fi, and 20 Mbps 4G LTE connection (AT\&T in the United States). Fig.~\ref{fig:eval1} represents the results of the experiments. As we can see from the evaluation, the similar address request time is consistent under different circumstances, and is around 2 seconds. Most importantly, as the number of addresses grows exponentially, we observe only a slight increase in delay. Specifically, \emph{While the number of addresses increased by 1,000,000\%, the delay increased only by 19.7\%}, which suggests that with larger database sizes the latency will still remain low.

Although it may be common to copy and paste text in under 2 seconds within a single window of a frequently used application, in the case of cryptocurrency transfer, the address will undoubtedly be copied from one application (e.g., web browser) and pasted into the hardware wallet client, which may also involve the application switching step to bring the wallet app to the foreground. Moreover, it is reasonable to assume that hardware wallet apps are not frequently used by most users because every cryptocurrency transfer incurs paying blockchain fees. Therefore, the workflow of the clipboard copy-paste cycle is likely to take more than 2 seconds on average. In our experiment, in which we repeated the workflow of the attack 12 times on a laptop (3 times for each wallet), the copy-paste delay exceeded 2 seconds each time (based on stopwatch measurements made by an observing assistant).

\textbf{Cooperative Account Transfer: }
We evaluated the delay of cooperative address transfer by conducting 5 experiments, each including 20 measurements (100 measurements total). In each experiment, we exponentially increased the number of addresses stored in each of the \emph{ClipperCloud} servers from $10^4$ to $10^8$. Then, for each of the 5 experiments, we measured the delay of transferring an Ethereum account from miner to cooperator in milliseconds, using two different \emph{ClipperCloud} servers (one in San Francisco, another one in New York). After that, we calculated the mean average and standard deviation of the 20 measurements for each experiment, and represented the results in Fig.~\ref{fig:eval2}.

As we can see from the results, the cooperative transfer time is consistent under different circumstances, and is in the order of 4 seconds. Most importantly, as the number of addresses grows exponentially, we observe only a slight increase in delay. Specifically, \emph{while the number of addresses increased by 1,000,000\%, the cooperative transfer delay increased only by 11.6\%}, which suggests that the complexity of the account transfer is $\mathcal{O}(1)$.

\subsection{Address Mining Performance}

In order to successfully conduct the \emph{EthClipper} attack, the attacker needs a large database of Ethereum accounts for address substitution, mined using a substantial compute power applied for a lengthy period of time. Therefore, it is crucial to evaluate the ability of an attacker to mine a \emph{ClipperCloud} database with desired parameters using a reasonable time and budget. In order to evaluate the performance of address mining by \emph{ClipperCloud}, we deploy four different server configurations. For the ease of reference, we give these configurations short code names: \emph{Azure}, \emph{DO-}, \emph{DO+}, and \emph{PC}. Below are details of these configurations:

\begin{itemize}
    \item \textbf{Azure:} Microsoft Azure H-Series HB60rs high-performance virtual machine with 60 CPUs, 223.52 Gb RAM, and 700 Gb of storage, running Ubuntu Server 20.04 LTS. The cost at the time of deployment (January 2021) was \$1,664.40/mo.
    \item \textbf{DO-:} DigitalOcean Basic Droplet with 1 vCPU, 1 Gb RAM and 25 Gb of storage, running Ubuntu 20.04 (LTS) x64. The cost at the time of deployment was \$5/mo.
    \item \textbf{DO+:} DigitalOcean CPU-Optimized Droplet with 32 CPUs, 64 Gb RAM, and 400 Gb of storage, running Ubuntu 20.04 (LTS) x64. The cost of the instance is \$640/mo.
    \item \textbf{PC:} Office PC with AMD Ryzen Threadripper x2950 CPU (16 cores, 32 threads), 70.6 Gb RAM, 1 Tb SSD storage, running Kubuntu 20.04 LTS.
\end{itemize}

\begin{figure}
    \centering
    \includegraphics[width=0.7\linewidth]{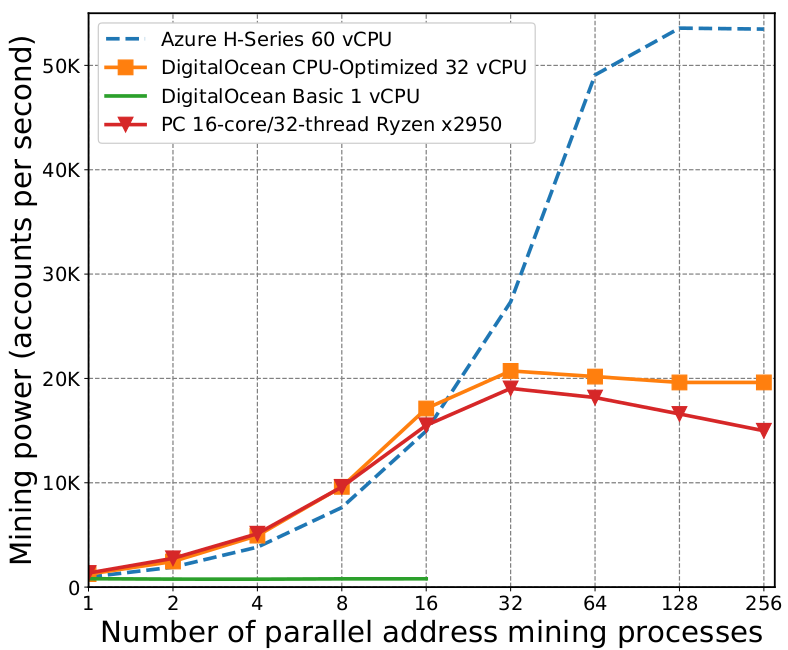}
    \caption{\textbf{\emph{ClipperCloud} address mining performance.}}
    \label{fig:miningperf}
\end{figure}

On each of the four configurations, we perform tests involving different number of simultaneous address mining processes: 1, 2, 4, 8, 16, 32, 64, 128, and 256. Each process mines and saves 100,000 random Ethereum accounts. For each test, we measure the time needed for all the threads to finish. Then, for each test, we calculate the mining performance measured in accounts per second for each of the server configurations. Fig.~\ref{fig:miningperf} shows the results of the experiments. The \emph{DO-} server hung each time when we attempted to run 32 simultaneous mining processes. Therefore, we were only able to gather partial data for it. All the servers, except \emph{DO-}, exhibit similar performance for 1, 2, 4, 8, and 16 simultaneous processes; however, the \emph{Azure} server shows a significant performance advantage with 32, 64, 128, and 256 simultaneous processes.

\begin{figure*}
\captionsetup[subfigure]{justification=centering}
\centering
\begin{subfigure}{.24\textwidth}
  \centering
  \includegraphics[height=1.2in]{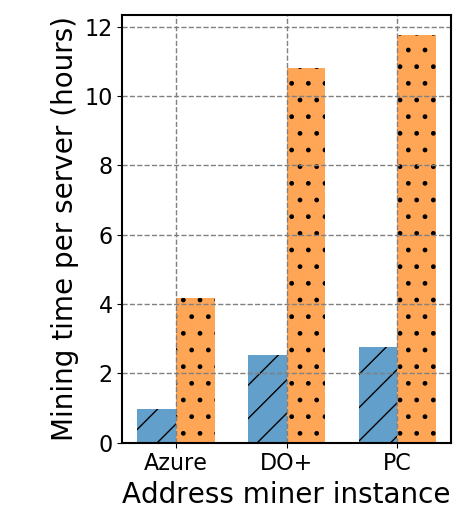}
  \caption{7 digits}
  \label{fig:time7}
\end{subfigure}
\begin{subfigure}{.24\textwidth}
  \centering
  \includegraphics[height=1.2in]{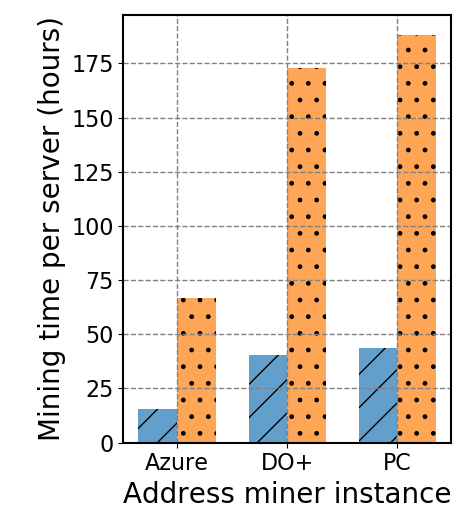}
  \caption{8 digits}
  \label{fig:time8}
\end{subfigure}
\begin{subfigure}{.24\textwidth}
  \centering
  \includegraphics[height=1.2in]{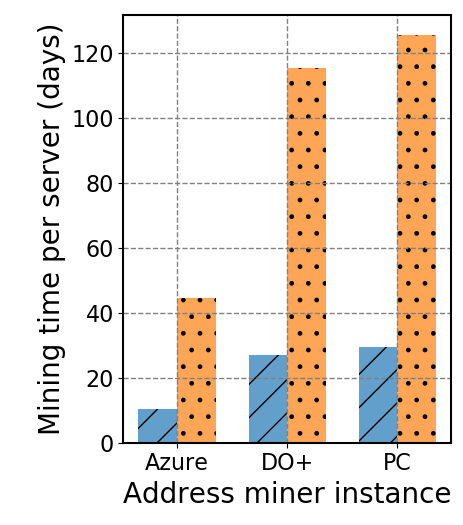}
  \caption{9 digits}
  \label{fig:time9}
\end{subfigure}
\begin{subfigure}{.24\textwidth}
  \centering
  \includegraphics[height=1.2in]{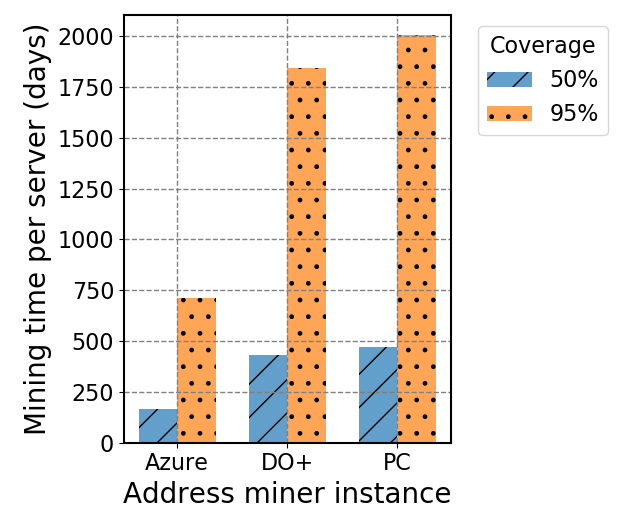}
  \caption{10 digits}
  \label{fig:time10}
\end{subfigure}

\caption{\textbf{Time needed to achieve the target address mining coverage for 7, 8, 9, and 10 matching digits.}
}
\label{fig:coverage}
\end{figure*}

Additionally, we evaluate how much time it would take for \emph{Azure}, \emph{DO+}, and \emph{PC} to mine 50\% and 95\% of the address database for 7, 8, 9, and 10 matching digits, with results shown in Fig.~\ref{fig:coverage} (please refer to \emph{Claim 1} for details of calculation of 95\% coverage). For the 50\% coverage, we use the same Taylor approximation with $\tau = 0.7$. First, we can see that, unsurprisingly, the \emph{Azure} deployment exhibits a significant performance advantage compared to \emph{DO+} and \emph{PC}. However, \emph{Azure} is also the most expensive deployment out of the three. Second, the performance difference between the \emph{DO+} and \emph{PC} deployments is insignificant, and given a sizable rental cost of \emph{DO+}, the use of retail PC may be the most economic option for an attacker, depending on available budget and other circumstances. 

Nevertheless, \emph{ClipperCloud} is suitable for a flexible variety of possible deployment scenarios. For example, a realistic and affordable scenario would be to use 5 office computers to mine a 50\% address coverage. The number of days to mine the required coverage for one 1 PC is 467.84, and if it is split between 5 computers, it would take $467.84 / 5 = 93.6 \approx 3\:\text{months}$. Essentially, it means that the attacker will be able to run the \emph{ClipperCloud} from home or office, statistically capable of replacing 50\% of incoming addresses with a 10-digit match, which is 25\% of all digits in an Ethereum address. Meanwhile, a small user study by Almutairi and Al-Megren~\cite{almutairi2019usability} demonstrated that 30\% of users of the KeepKey wallet failed to recognize the substitution of a Bitcoin address with 20\% of matching symbols. Since \emph{EthClipper} is an opportunistic attack, a success rate around 30\% is capable to yield a substantial gain for the attacker. In conclusion, the \emph{EthClipper} attack is a realistic attack which can be launched by an attacker with relatively limited resources. 
%Since an Ethereum address has 40 digits total, the 10 digit match accounts for 25\% of the entire address length.

\subsection{Opinions from Manufacturers of Hardware Wallets}
For responsible disclosure, we contacted all the manufacturers of the wallets used in this research, and all of them confirmed the potential danger of the attack. Specifically the security representative from ShapeShift stated, ``\emph{[...] it would likely impact KeepKey users since in my experience, you are right: most users either verify the first/last characters or none at all.}'' The security representative from SatoshiLabs s.r.o. stated, ``\emph{It's quite obvious from the description how the attack works [...]}'' The head of security research at Ledger SAS said, ``\emph{The attack you described is a problem we already discussed, and we did not find a satisfactory solution to tackle it. We would be happy to collaborate with you in order to develop defenses against it.}'' Following the responses, we are in the process of discussing a collaborative defense solution against the attack.

\section{Security Recommendations and Defense}
In this section, we discuss two categories of measures that can be used to prevent the \emph{EthClipper} attack: adherence to security recommendations and automated defense against the attack.

\subsection{Security Recommendations}

\noindent\textbf{Recommendation 1: Resist confirmation bias.}
\emph{EthClipper} is a hybrid attack with a substantial social engineering component, which means that its success largely depends upon the ability of the attacker to exploit the human cognitive bias of the user of a hardware wallet. Specifically, the attack relies on the confirmation bias that forces the user to conclude that the actual recipient address matches the intended one, based on a partial reading.
However, a proper verification of the entire address by the user, prior to sending funds to it, is sufficient to reveal the address substitution. Therefore, a disciplined verification of the entire address by the user would deliver a reliable defense against the \emph{EthClipper} attack.

\noindent\textbf{Recommendation 2: Pay attention to EIP-55 checksums.}
Ethereum clients often use address checksums, also known as \emph{EIP-55 checksums}, which are encoded in the addresses via selective capitalization of certain hexadecimal letters. These checksums are primarily designed for software clients to detect typos in hand-typed addresses; however, they can also be useful for uncovering an \emph{EthClipper} attack. Although an EIP-55 capitalization can be falsified~\cite{ivanov2021targeting}, it would incur a significant computation overhead for the \emph{ClipperCloud} address miner\footnote{The probability of EIP-55 checksum collision is $\approx$0.0139\%~\cite{antonopoulos2018mastering}.}, rendering the creation of the address database impossible within a reasonable time frame. Consequently, the address substituted by \emph{EthClipper} would likely have different capitalization than the original one. Therefore, when verifying the correctness of Ethereum addresses, we recommend to pay attention to the capitalization of their hexadecimal letters.

\subsection{Automated Defense}
In the spirit of reproducible and open research, we intend to make the source code of the \emph{EthClipper} stack published after the defense is developed and incorporated by the wallet manufacturers. The malware component of our stack can be reused to implement a resident program that issues notifications or sound alarms each time a new Ethereum address is added to the operating system clipboard. Moreover, we are in the process of submitting recommendations to all the vendors of hardware wallets to incorporate the clipboard monitoring components into their desktop client software. Specifically, a system alert issued by such a component upon detection of an Ethereum address in the clipboard would effectively prevent the event of address substitution to be unnoticed, because the user will always see a system notification whenever an address appears or replaced in the clipboard.

\section{Related Work}\label{sec:related-work}

The research related to hardware wallets mostly focuses on hardware vulnerabilities and feature enhancements.
Guri \textit{et al.}~\cite{guri2018beatcoin} demonstrate a technique that allows an attacker to exfiltrate private keys from a hardware wallet by installing a malware directly on the wallet's firmware. Gutoski \textit{et al.} \cite{gutoski2015hierarchical} show that the hierarchical deterministic (HD) wallet design, used in all popular hardware wallets, could reveal all the private keys in the hierarchy if only one of the private keys is leaked; this research further proposes a new design of an HD wallet that allows to avoid such key co-dependency. Several studies in wireless sensing~\cite{li2021deep} demonstrate the ability to steal passcodes from personal devices, possibly including hardware wallets. The above adversarial scenarios, however, assume that either the attacker has a physical access to the hardware wallet, or there is a partial leak of wallet credentials --- intuitively, both the scenarios are highly unlikely within the context of the \emph{EthClipper} attack, which zeroes in on the adversarial actions that the real world attackers have been using successfully for decades, i.e., malware infestation of user computers and social engineering. Moreover, Datko et al.~\cite{datko2017breaking} demonstrate how the firmware of some hardware wallets can be attacked to steal the user PIN code. San Pedro et al.~\cite{san2019side} explore side-channel attacks that allow to extract PIN codes and private keys from \emph{Trezor One} hardware wallet --- although the vulnerability has been timely patched by the manufacturer, it demonstrates that the hardware and firmware components of hardware wallets can also be attacked. Gkaniatsou et al.~\cite{gkaniatsou2017low} show how the low-level local communication protocol between the client software and the hardware wallet can be used for side-channel attacks. Nevertheless, 
%while breaking the air-gap protection of hardware crypto wallets is unrelated to \emph{EthClipper}, 
fixing these hardware vulnerabilities does not make hardware wallets less susceptible to the \emph{EthClipper} attack.

\section{Conclusion}\label{sec:conclusion}
Hardware crypto wallets are relatively expensive and popular among the users who own large amounts of cryptocurrency. These devices promise the protection of the stored funds even in the event when the attacker gains full control over the victim's computer, including the malware invasion scenario. However, in this work we demonstrated that it is possible to compromise the air-gapped security of a hardware wallet and fool its owner into confirming a malicious transaction, even without jeopardizing the integrity of the wallet itself. Our \emph{EthClipper} attack, which is confirmed to be potentially dangerous by the manufacturers of three leading hardware wallet firms, not only falsifies the input to the hardware wallet, but it also crafts the address in a way that allows to circumvent the transaction verification procedure. Our evaluation confirms that the attack can be carried out with a limited budget on a retail equipment. As hardware wallets continue populating the market, we anticipate a growing number of opportunistic social engineering attack attempts on these wallets, and we believe that our paper will raise the vigilance about such attacks. At the time of writing, there is no affiliation or sponsorship, current or arranged, between the authors of this work and the manufacturers of the hardware wallets used in this research.

\section*{Acknowledgement}
We would like to thank anonymous reviewers for their valuable feedback on our work.

\bibliographystyle{IEEEtran}
% Generated by IEEEtran.bst, version: 1.14 (2015/08/26)

\end{document}